\documentstyle[aps,preprint,tighten,epsf,rotate,floats]{revtex}
\newbox\rotbox

\def\slash#1{\rlap/{#1}}
\def\MeV{\nobreak\,\mbox{MeV}}
\def\GeV{\nobreak\,\mbox{GeV}}

\def\beq{\begin{equation}}
\def\eeq{\end{equation}}
\def\mytoday#1{{}\ifcase\month\or
January\or February\or March\or April\or May\or June\or
July\or August\or September\or October\or November\or December\fi
 \space \number\year}

\begin{document}
\preprint{\vbox{Submitted to {\it Phys.\ Lett.\ B }\hfill
DOE/ER/40762--048\\
                \null\hfill UMPP \#95--035}}
\vskip 1 cm
\title{The Electromagnetic Pion Form Factor and Instantons}
\author{Hilmar Forkel\thanks{Address after Nov. 1, 1994:
European Centre for Theoretical Studies in Nuclear Physics and
Related Areas, Villa Tambosi, Strada delle Tabarelle 286, I-38050
Villazzano, Italy} and Marina Nielsen%
\thanks{Permanent address:
Instituto de F\'\i sica, Universidade de S\~ao Paulo, 01498 -
SP- Brazil.} }
\address{Department of Physics and Center for Theoretical
Physics } \address{University of Maryland, College Park, MD
\ 20742-4111 (U.S.A.)}
\date{\mytoday}
\maketitle

\vskip 1 cm
\begin{abstract}

We calculate the electromagnetic pion form factor at intermediate
space-like momentum transfer from the QCD sum rule for the
correlation function of two {\it pseudoscalar } interpolating fields
and the electromagnetic current. This correlator receives essential
contributions from direct ({\it i.e.} small-scale) instantons, which
we evaluate under the assumption of an instanton size distribution
consistent with instanton liquid and lattice simulations. The
resulting form factor is in good agreement both with the sum rule
based on the axial-current correlator and with experiment.

\end{abstract}
\newpage

%%%%%%%%%%%%%%%%%%%%%%%%%%%%%%%%%%%%%%%%%%%%%%%%%%%%%%%%%%%%%%%%%%%%%%%%
%123456789 123456789 123456789 123456789 123456789 123456789 123456789 12

A central goal of strong-interaction physics is the
understanding of hadrons on the basis of quantum chromodynamics (QCD).
It is very unlikely that this goal can be reached without a thorough
understanding of the QCD vacuum.  At present, however, direct links
between observed hadron properties and the vacuum structure are
still rare and rely mostly on extensive numerical simulations.
The aim of this letter is to study one such link -- between direct
instantons, {\it i.e.} small-scale topological vacuum fields, and
electromagnetic pion properties -- analytically in the framework of
a QCD sum rule \cite{svz,rry}.

Due to the Goldstone nature of the pion, sum rule calculations
of pionic properties can be based on two in principle (but not in
practice!) equally suitable sets of correlation functions,
corresponding to the use of pseudoscalar or axial vector interpolating
fields. The pion couples strongly to both of these source currents
and thus contributes to the correlators in both channels.

The pseudoscalar channel has some principal advantages for sum rule
calculations. The accuracy of the standard pole-continuum
parametrization for the corresponding spectral functions profits
from the almost complete dominance of the pion in the low-mass region
\cite{nsvz,shurev}. Furthermore, correlators involving the pseudoscalar
interpolators have a simpler tensor structure. This simplifies in
particular the calculation of three-point functions.

All existing sum rule applications in the pion channel (with the
exception of ref. \cite{shu1}, see below), however, are based on
axial vector current correlators \cite{svz,rry,iof3,nera}. The use
of the pseudoscalar interpolating field has been avoided, since it
is known to receive essential contributions from direct instantons
\cite{nsvz}. Like instanton contributions in general, they could
initially not be estimated reliably, due to insufficient
knowledge of the instanton size distribution in the vacuum. The
attempt of an {\it ab initio} description in the dilute gas
approximation \cite{CDG}, in particular, failed for all but the
smallest instanton sizes due to infrared problems with large
instantons. The same problems were encountered in the attempt to
supplement the conventional operator product expansion (OPE) in
QCD sum rules with direct instanton contributions and led to the
preference for the axial vector correlators.

In the last decade, however, mutually consistent information about
the instanton size distribution has been gathered from new sources,
including phenomenological estimates \cite{shu3}, variational studies
\cite{DiaPe84} and numerical simulations of instanton liquid
models \cite{ShuVer90}, as well as lattice calculations \cite{ChuHua92}.
The bulk features of the density $n(\rho)$ of instantons with size
$\rho$ in the vacuum, which emerged from these studies, can be
summarized in a simple parametrization \cite{shu3},
\beq
n(\rho) = \bar{n} \, \delta(\rho-\bar{\rho}) \; , \label{n}
\eeq
with the average (anti-) instanton density and size fixed at
\beq
\bar{n} =  \frac12 \, {\rm fm}^{-4}, \qquad \qquad \quad \bar{\rho} =
\frac13 {\rm fm} \;. \label{instpar}
\eeq
This form provides a reasonable approximation to the sharply
peaked gaussian distribution found in ref. \cite{ShuVer90} and
should be sufficiently accurate for our purpose.

Armed with the bulk features of the instanton density, we can now
try to turn the former vice of the pseudoscalar interpolating field --
its strong coupling to instantons -- into a virtue, by using
pseudoscalar correlators as a tool for ``instanton diagnostics''.
Linked by the corresponding sum rules to observable hadron properties,
these correlators provide a theoretical laboratory both for the study
of instanton effects at short and intermediate distances and for
tests of the approximations used to calculate them. They also allow
an independent check of the instanton distribution (\ref{n}).

With this motivation in mind, we calculate in the present paper
the electromagnetic pion form factor from a QCD sum rule based on
pseudoscalar interpolating fields. The instanton contributions are
evaluated semiclassically in the zero-mode approximation. Our
investigation complements similar studies of the nucleon \cite{for93}
and pion \cite{shu1} mass sum rules.

The sum rule techniques for the calculation of hadron form factors
at intermediate momentum transfers were developed in refs.
{}~\cite{iof3,nera}. Their application to the pion form factor, based
on axial vector interpolators, leads to a rather good agreement
with experiment (in the momentum transfer region $0.5 < Q^2 < 3.0 \,
\GeV^2$) \cite{iof3,nera}. Our calculation will thus also provide a
consistency check between the axial vector and pseudoscalar sum
rules.

To derive the sum rule for the pion form factor, we consider the
three-point correlation function of two pseudoscalar currents
$j_5(x)= \overline d(x) i\gamma_5 u(x)$ and the electromagnetic
current  $j_\mu^{el}(x)=e_u \, \overline u(x) \gamma_\mu u(x) +
e_d \, \overline d(x)\gamma_\mu d(x)$,
\beq
\Gamma_\mu(p,p';q) = -\int d^4{x}\int d^4{y}e^{i(p'\cdot x-q\cdot y)}
\langle 0|\, T\, j_5^\dagger(x) \, j_\mu^{el}(y) \, j_5(0) \, |0\rangle \ ,
\label{corr}
\eeq
with $q=p'- p$.  $\Gamma_\mu$ can be decomposed into two independent
Lorentz vector structures,
\beq
\Gamma_\mu(p,p';q) = \Gamma_1(p^2,{p'}^2,q^2)\, (p'+p)_\mu +
\Gamma_2(p^2,{p'}^2,q^2)\, q_\mu \; .
\label{str}
\eeq
The invariant amplitudes $\Gamma_{1,2} $ have a double dispersion
representation of the form
\beq
\Gamma_i(p^2,{p'}^2,q^2) = {1\over\pi^2}\int_0^\infty ds \int_0^\infty
ds' {\rho_i(s,s',q^2)\over (s-p^2)(s'-{p'}^2)} + ... \; ,
\label{dire}
\eeq
where the ellipsis represents subtraction polynomials in $p^2$ and
${p'}^2$, which will vanish under the Borel transform to be applied
later.

Since the pseudoscalar current has a nonvanishing matrix element
between the vacuum and the one-pion state,
\beq
\langle 0|\overline d i \gamma_5 u |\pi^+\rangle= {\sqrt{2}
f_\pi K } \; , \qquad  K = \frac{m_\pi^2}{m_u+m_d} \label{overl}
\eeq
($f_\pi=93\MeV$ is the pion decay constant and $m_{u,d}$ are the
up and down current quark masses), the pion contribution to the
spectral density is of the form
\beq
\langle 0| j_5^\dagger |\pi(p')\rangle \langle\pi(p')|j_\mu^{el}|\pi(p)
\rangle \langle\pi(p)| j_5|0\rangle = 2 \,f_\pi^2\, K^2 \,Q_\pi \,
F_\pi(Q^2) \, (p'+p)_\mu  \ ,
\label{mae}
\eeq
where $Q^2=-q^2$, $Q_\pi$ is the charge of the pion and $F_\pi(Q^2)$
is the form factor to be calculated. The latter is contained in the
amplitude $\Gamma_1$, on which we will concentrate in the following.
For the continuum contribution we adopt the standard form of refs.
\cite{iof3,nera}, which completes our parametrization of the spectral
density $\rho_1$:
\beq
\rho_1(s,s',Q^2) = 2 \,f_\pi^2 \, K^2 \,  Q_\pi \, F_\pi(Q^2)\,
\, \delta(s) \,\delta(s') + \theta(s + s'- s_0) \, \rho_0(s,s',Q^2)
\label{specd}.
\eeq
($\rho_0(s,s',Q^2)$ is the free-quark spectral function and we
neglected the small pion mass.) The continuum threshold $s_0$ defines
a triangular region in the $(s,s')$ plane and is related to the
threshold $s_1$ of the corresponding pseudoscalar two-point correlator
as $s_0 \simeq 1.5 \, s_1$ (see ref. \cite{iof3} for details).

An alternative continuum ansatz with a quadratic integration region
bounded by $s_1$ was also considered in \cite{iof3} for the
axial-vector three-point function. Similar to ref. \cite{iof3} we
find almost no difference between the two parametrizations in the
resulting form factor \cite{big}.

The leading instanton (and anti-instanton) contributions to $\Gamma_\mu$
are calculated in semi-classical approximation, which amounts to
evaluating the correlator (\ref{corr}) in the background of the
(anti-) instanton fields and to averaging their collective coordinates
over the corresponding vacuum distributions. We will outline only the
essential features of this procedure and refer to the literature
\cite{shu1,shu3,for93} for more details.

Since the sum rule probes the correlation function mainly at distances
$x,y \simeq 0.2 \, {\rm fm}$, which are small compared to the average
inter-instanton separation $\bar{R} \simeq 1 \, {\rm fm}$, the
correlated contributions from two or more instantons should be small.
We are thus led to treat only the effects of the (anti-) instanton
closest to $x$ and $y$ explicitly and to include effects of interactions
with other vacuum fields (including other instantons) at the mean-field
level \cite{svz2}.

To obtain the explicit instanton contribution, we recall that the
quark spectrum in the background of an (anti-) instanton contains
one zero-mode state per flavor \cite{thooft},
\beq
\psi_0^\pm(x) = \frac{\rho}{\pi} \frac{1 \pm \gamma_5}{(r^2 +
\rho^2)^{3/2}} \, \frac{\slash{r}}{r} \, U, \label{zm}
\eeq
with $r = x-z$, where $z$ is the instanton position.
(The spin-color matrix $U$ satisfies $(\vec{\sigma} + \vec{\tau})
\, U = 0$.) Up to corrections of order $m^* \bar{\rho}$ (see below)
from higher-lying states, quarks in these two zero-modes dominate the
instanton contribution to the correlator (\ref{corr}). Their
contribution is evaluated by inserting the explicit expression
(\ref{zm}) into (\ref{corr}) and by approximating quarks propagating
in the higher-lying (continuum) states as non-interacting.

For quarks propagating in zero-mode states the interaction with the
other vacuum fields leads at the mean-field level to the generation
of an effective quark mass $m^*(\rho)= -{2\over3}\pi^2\rho^2\langle
\overline{q}q\rangle$ \cite{svz2}, which replaces the current mass
in the zero-mode propagator. Taking this effect into account, we
obtain (after a Wick-rotation to euclidean space-time) the following
instanton contribution to the correlator (\ref{corr}):
\begin{eqnarray}
\Gamma_\mu(p,p';q) &=& -{4Q_\pi\over \pi^6} {\bar{n} \,
\bar{\rho}^4\over {m^*}^2(\bar{\rho})}
\int d^4{x}\int d^4{y}\int d^4{z}\, e^{ip'\cdot x}e^{-iq\cdot y}
e^{i(p'-q)\cdot z}\times
\nonumber\\*[7.2pt]
& &\left[ {y^2 \, (y+z)_\mu -(y+z)^2 \, y_\mu \over
(x^2+\bar{\rho}^2)^3y(y^2+\bar{\rho}^2)^{3/2}(y+z)^4z(z^2+
\bar{\rho}^2)^{3/2}} + \left(\begin{array}{c} z_\mu \leftrightarrow
x_\mu   \\ y_\mu \rightarrow - y_\mu \end{array} \right) \right] \; ,
\label{inscorr}
\end{eqnarray}
In this expression we summed over instanton and anti-instanton
parts and integrated over their positions. Since we deal with
a gauge-invariant correlator, the averaging over the color
orientations is trivial, and the average over the remaining
collective coordinate, the instanton size $\rho$, was weighted
with the distribution (\ref{n}).

After taking the standard double Borel transform \cite{iof3}, both in
$p^2$ and $p'^2$, of (\ref{inscorr}) (details of the rather lengthy
calculation will be given in \cite{big}), we obtain
\beq
\Gamma_1^{(in)} (Q^2, M^2) = -Q_\pi \, {\bar{n} \, M^2 \over
{m^*}^2(\bar{\rho})}
\, I_{inst}(\bar{\rho}^2 Q^2, \bar{\rho}^2 M^2)\; ,\label{insfi}
\eeq
($M$ is the Borel mass) in terms of the dimensionless integral
\begin{eqnarray}
I_{inst} (\tilde{z}^2, z^2)&=&\int_0^\infty d\alpha \int_0^{z^{-2}}
d\epsilon \, e^{-\alpha \tilde{z}^2} e^{-(\alpha'+\gamma')}
e^{-z^2\over 4(1-\epsilon z^2)}{\alpha\epsilon\over A^4(1-\epsilon
z^2)} \times
\nonumber\\*[7.2pt]
& &\left\{H(\alpha')H(\gamma')\left[{\alpha+\epsilon\over z^2}
\left(\alpha\  \tilde{z}^2 +{\epsilon z^2\over 16}{z^2-8(1-\epsilon
z^2)\over(1-\epsilon z^2)^2} -3\right)\right.\right. -
\nonumber\\*[7.2pt]
& &\left.\left. {2\alpha\epsilon\over A}(2\alpha\epsilon-A) -
{\alpha^3\epsilon\over4z^2A^2}\right] - {\alpha\epsilon(\alpha+
\epsilon)\over z^2A}I_1(\alpha')H(\gamma') + {\alpha^2\epsilon^2
\over A}H(\alpha')I_1(\gamma')\right\}\; ,
\label{insint}
\end{eqnarray}
where $A = {\alpha+\epsilon\over z^2} +
\alpha\epsilon $, $\alpha'= {\alpha\over 8A}$, $\gamma'={1\over 8z^2A}$,
and $H(x)= I_1(x)- I_0(x)$ is defined in terms of the modified Bessel
functions $I_n(x)$.

The lowest-order perturbative contribution to $\Gamma_1$ was evaluated
in ref. \cite{iof3}\footnote{Note that the corresponding expression
in \cite{iof3} contains a misprint.},
\beq
\Gamma_1^{(pert)} (Q^2, M^2) = {3Q^2M^2\over 16\pi^2}
I_{pert}(Q^2, M^2)\; ,\label{pert}
\eeq
where the continuum contributions from the parametrization
(\ref{specd}) are subtracted in the integral
\beq
I_{pert}(Q^2,M^2) = \int_0^{s_0/M^2} dx\, e^{-x}\int_0^x dy\,
{x^2-y^2\over(Q^4/M^4+2xQ^2/M^2+y^2)^{3/2}}.
\eeq
As already noted in ref.\cite{iof3}, the power corrections to
$\Gamma_1^{(pert)}$ are small in the $Q^2$ range under
consideration\footnote{See also the related discussion for the
two-point correlator in \cite{nsvz}.} and will be neglected in
this letter. A more complete analysis of the OPE can be found
in \cite{big}.

Adding instanton (\ref{insfi}) and perturbative (\ref{pert})
contributions and equating them to the Borel transformed dispersion
representation (\ref{dire}) leads to the sum rule
\begin{eqnarray}
2 \, f_\pi^2 \, K^2 \, F_\pi(Q^2) &=& \left[{3Q^2M^2\over 16\pi^2}
I_{pert}(Q^2,M^2) - {\bar{n}\, M^2 \over {m^*}^2(\bar{\rho})}
I_{inst} (\bar{\rho}^2 Q^2, \bar{\rho}^2 M^2)\right]L^{-8/9}\; .
\label{fpi}
\end{eqnarray}
The factor $L^{-8/9}$, with $L=\ln(M^2/\Lambda_{\rm QCD}^2)/
\ln(\mu^2/\Lambda_{\rm QCD}^2)$, accounts for the scaling behavior
of the three-point correlator due to the anomalous dimension of the
pseudoscalar currents, and sets their renormalization point to
$\mu=500\MeV$. (We use the value $\Lambda_{\rm QCD}=150\MeV$ for
the QCD scale parameter.)

Let us now fix the remaining parameters in the sum rule. As already
mentioned, the continuum threshold $s_0$ can be related to the
corresponding threshold $s_1$ of the pseudoscalar two-point
correlator, $s_0 \simeq 1.5 \, s_1$. The analysis of \cite{shu1}
(see also ref. \cite{nsvz}) finds  $s_1\simeq 2.0 \, \GeV^2$ and we
will thus fix the threshold at $s_0 = 3.0 \, \GeV^2$. The relatively
large separation of the continuum from the lowest resonance in the
pseudoscalar channel (about twice of that in the vector and axial
vector channels) is clearly favorable for the sum rule analysis,
since it improves the accuracy of the parametrization (\ref{specd}).

Due to uncertainties in the determination of the light quark
masses, the phenomenological value of the (scale-dependent) mass
parameter $K(\Lambda)$, defined in eq. (\ref{overl}), is at present not
accurately known. From the current bounds on the up and down quark
masses \cite{gasser} and with $m_\pi=138\MeV$ we obtain $1\leq K (1 \,
\GeV) \leq 2 \, \GeV$. In order to allow for a direct comparison
with the instanton analysis of the two-point correlator
\cite{shu1}, which takes $K = 0.7 \, \GeV$, we fix $K$ at the lower limit
of the phenomenologically acceptable range, $K = 1.0 \,\GeV$.
The effective mass $m^*$, finally, is determined as in
\cite{shu1,for93} from the self-consistency relation \cite{CDG} for
the quark condensate, $\langle \overline q  q\rangle = -2\, \bar{n}
/m^*(\bar{\rho})$.

Let us now proceed to the quantitative analysis of the sum rule
(\ref{fpi}). Figure 1 shows the $M^2$ dependence of both the
perturbative and the instanton contributions to the pion form factor
at fixed $Q^2=1 \,\GeV^2$. Two features of the instanton part are
immediately apparent: First, and as expected, it is the {\it dominant}
contribution, about a factor of two larger than the perturbative
part. Secondly, it improves the stability of the sum rule, {\it i.e.}
it reduces the $M^2$ dependence of the form factor, and a
``stability plateau'' begins to develop for $M^2 > 1.2\, \GeV^2$.
The same qualitative behavior is found for all values of $Q^2
\geq 0.5\, \GeV^2$.

In order to determine the $Q^2$ dependence of the form factor, we
follow the procedure of ref. \cite{iof3} and evaluate the sum
rule for different values of $Q^2$ at a fixed value of the Borel
mass, $M^2=1.2\GeV^2$, postponing a more accurate analysis to
ref.\cite{big}. The resulting form factor is shown in fig.2 and
compared with the experimental data of \cite{exp} at the space-like
momentum transfers accessible within our approach\footnote{The
perturbative part dominates the instanton contributions at $Q^2
\rightarrow \infty$ and shows the expected $(Q^2)^{-2}$ behavior.
The so far neglected power corrections, however, will eventually
blow up \cite{big}, thereby limiting the applicability of the sum
rule at large $Q^2 $.}.
The agreement is clearly satisfactory and at least as good as the
one from the axial vector sum rule \cite{iof3} (dashed line), which
we show for comparison.
(A slightly better agreement between the axial vector sum rule
result and the data was obtained in ref. \cite{nera} by extracting
the $Q^2$ dependence of $F_\pi$ at $M^2=1.8 \,\GeV^2$.)

The somewhat better fit of the pseudoscalar sum rule result at low
$Q^2$ (where also the data become more reliable) is perhaps not
accidental. The result of the axial-vector based sum rule relies
exclusively on the OPE, which breaks down at small $Q^2 $,
since physics of longer distance scales begins to determine the
behavior of the three-point function. The instanton contribution,
in contrast, which dominates our result, could optimistically
remain reliable up to distances not too far below the inter--instanton
separation, corresponding to $Q^2 \simeq 0.1 - 0.2 \,\GeV^2$.
Indications supporting this expectation have been found in an
analogous study of the nucleon correlation functions \cite{for94}.
The applicabilty of the sum rule at small $Q^2$ would then be
mainly limited by the unphysical singularity in the perturbative
contribution (\ref{pert}).

To conclude, we view the above results, and in particular the
stability of the pseudoscalar sum rule and its agreement with
phenomenology, as further support for the importance of the
instanton component in the QCD vacuum, for the bulk features
of their distribution as given in eq. (\ref{instpar}), and for
the semiclassical estimate of their effects in hadron correlators
at intermediate distances.

%%%%%%%%%%%%%%%%%%%%%%%%%%%%%%%%%%%%%%%%%%%%%%%%%%%%%%%%%%%%%
%123456789 123456789 123456789 123456789 123456789 123456789 123456789 12

%\newpage

H.F. acknowledges support from the Department of Energy under
Grant No.\ DE--FG05--93ER--40762 and M.N. acknowledges support
from FAPESP  Brazil.

%%%%%%%%%%%%%%%%%%%%%%%%%%%%%%%%%%%%%%%%%%%%%%%%%%%%%%%%%%%%%%%%%%%%%%

% \end{thebibliography}
%\eject
\newpage

\begin{figure}[bht]
\caption{Borel mass dependence of the pion form factor from
Eq.~(\protect{\ref{fpi}}) at $Q^2 = 1\, {\rm GeV}^2$ (solid curve).
The dashed and dotted curves correspond to the instanton and the
perturbative contributions, respectively.}
\label{fig1}
\end{figure}

\begin{figure}[bht]
\caption{The electromagnetic pion from factor from
Eq.~(\protect{\ref{fpi}}) (solid line) at $M^2=1.2 \GeV^2$. The
dashed line shows for comparison the result of the axial vector
sum rule ~\protect{\cite{iof3}}. The experimental data are taken
from ref.~\protect{\cite{exp}}.}
\label{fig2}
\end{figure}

\end{document}